# Switching the Optical Chirality in Magneto-plasmonic Metasurfaces Using Applied Magnetic Fields

Jun Qin,[1] Longjiang Deng,[1] Tongtong Kang,[1] Lixia Nie,[1] Huayu Feng,[2] Huili Wang,[1] Run Yang,[1] Xiao Liang,[1,5] Tingting Tang,[5] Chaoyang Li,[4] Hanbin Wang,[3] Yi Luo,[3] Gaspar Armelles[2] and Lei Bi[1,*]

[1]National Engineering Research Center of Electromagnetic Radiation Control Materials, University of Electronic Science and Technology of China, Chengdu 610054, China

[2]Instituto de Micro y Nanotecnologíá (INM-CNM-CSIC), Isaac Newton 8, PTM, E-28760, Tres Cantos, Madrid, Spain

[3]Microsystem and Terahertz Research Center, China Academy of Engineering Physics, Chengdu, 610200, China

[4]Hainan University, No. 58, Renmin Avenue, Haikou, Hainan Province, 570228, P. R. China

[5]College of Optoelectronic Technology, Chengdu University of Information Technology Chengdu, 610225, China

*Corresponding author: bilei@uestc.edu.cn

**ABSTRACT**

Chiral nanophotonic devices are promising candidates for chiral molecules sensing, polarization diverse nanophotonics and display technologies. Active chiral nanophotonic devices, where the optical chirality can be controlled by an external stimulus has triggered great research interest. However, efficient modulation of the optical chirality has been challenging. Here, we demonstrate switching of the extrinsic chirality by applied magnetic fields in a magneto-plasmonic metasurface device based on a magneto-optical oxide material, $Ce_1Y_2Fe_5O_{12}$ (Ce:YIG). Thanks to the low optical loss and strong magneto-optical effect of Ce:YIG, we experimentally demonstrated a giant and continuous far-field circular dichroism (CD) modulation by applied magnetic fields from -0.65° to +1.9° at 950 nm wavelength under glancing incident conditions. The far field CD modulation is due to both magneto-optical circular dichroism and near-field modulation of the superchiral fields by applied magnetic fields. Finally, we demonstrate magnetic field tunable chiral imaging in millimeter-scale magneto-plasmonic metasurfaces fabricated using self-assembly. Our results provide a new way for achieving planar integrated, large-scale and active chiral metasurfaces for polarization diverse nanophotonics.

Chirality describes the symmetry property of a structure, that its mirror image cannot be superimposed with itself through translation and rotation operations, like our two hands. The chirality of biomolecules is universal in our living body, such as amino acid and proteins, which has significance in biomolecules recognition.[1,2] However, the chiroptical signal of chiral biomolecules is very weak. Recently, chiral plasmonic[3,4] and all dielectric structures[5,6] with large chiroptical response have attracted great research interest. Benefitted from advanced nano-fabrication technologies, 3D or planar chiroptical nanostructures, such as helices,[7,8] shurikens,[9] gammadions,[6,10] and twisted split-rings[11] have been fabricated. On the other hand, extrinsic chirality can also be observed in achiral photonic nanostructures under obliquely incidence conditions. The extrinsic chirality is originated from asymmetric distributions of electromagnetic fields, *i.e.* the electromagnetic near field distribution is chiral.[12] Nanophotonic structures such as nanoholes,[12] squares[13] and split ring resonators[14,15] showed large extrinsic chirality. For instance, Ben *et al.* demonstrated a ~4 times stronger optical chirality in achiral periodic nanoholes compared to the gammadion structure.[12] Recently, Abraham *et. al.* experimentally demonstrated that the achiral nanohole structures can also show superchiral near-fields even at perpendicular incidence conditions.[16] In that case, the far-field circular dichroism (CD) background signals from the plasmonic structures is eliminated, improving the sensitivity for chiral molecule sensing applications. These reports demonstrate a promising potential of utilizing extrinsic chirality for integrated biomedical sensing and polarization diverse photonic devices.

Recently, active chiral metamaterials have attracted great research interest. Unlike biomolecules showing fixed optical chirality, the chiroptical response of an artificial nanostructure can be switched by an external stimulus. Several methods have been demonstrated for chiroptical switching, such as phase-change materials ($VO_2$, $Ge_3Se_2Te_6$),[18,19] DNA origami,[20,21] mechanical deformation,[22] chemical reactions[23] and magneto-optical effects.[10,24-26] Among these methods, active control of the optical chirality using magneto-optical effects have attracted particular strong research interest.[27,28] Compared to other mechanisms, modulation of the optical chirality using magnetic fields shows advantages of high speed, low power consumption, continuous tunablity and ease for integration.[10,26] For instance, a far field CD modulation amplitude up to 150% is observed in an Au-Au-Ni trimer chiral plasmonic device.[26] However, the weak magneto-optical effect and strong optical absorption in ferromagnetic metals limited the modulation efficiency. The magnetic field induced CD modulation is usually much weaker than the structural CD, resulting in small modulation amplitudes of the far field CD. Therefore, efficient control of the optical chirality in magneto-optical chiroptical metasurfaces is yet to be demonstrated.

Here, we demonstrate continuous switching of the extrinsic chirality in a magneto-optical metasurface by applied magnetic fields. In particular, we observed a sign reversal for the far field CD upon reversing the magnetic field with a large CD modulation amplitude up to ~2.5°, which is more than one order higher compared to previous reports.[10,24,26] By using self-assembly fabrication methods, we also demonstrate large scale integration and chiral imaging properties of such structures. The large optical chirality modulation in such structures is fundamentally due to the strong magneto-optical effect and low optical loss of Ce:YIG thin films, therefore demonstrating their promising potential for active chiroptical metasurface applications.

## RESULTS AND DISCUSSION

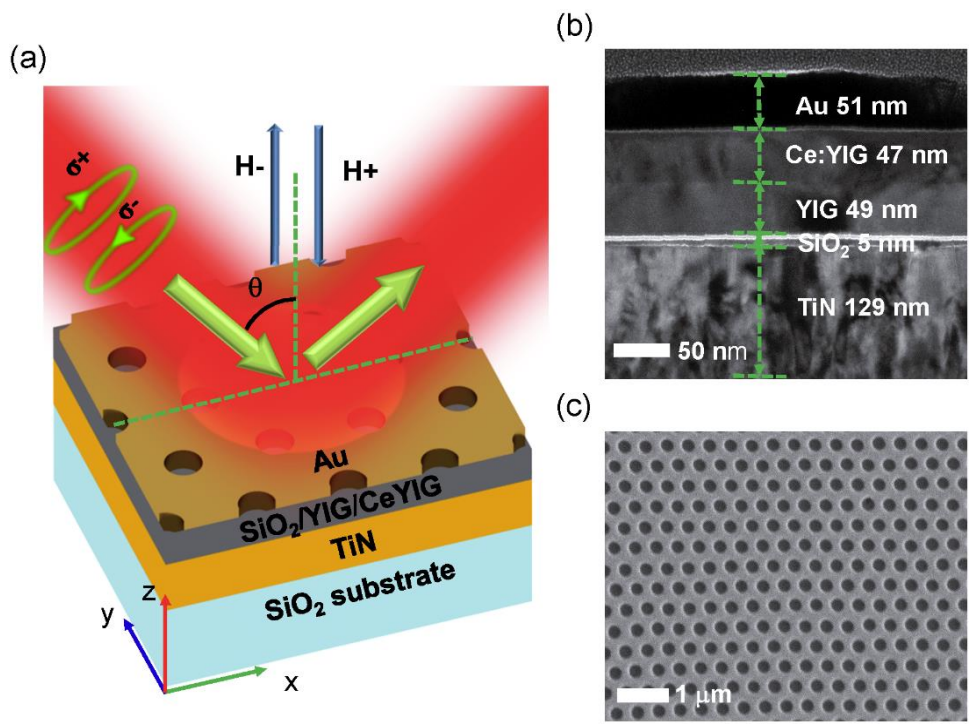

**Figure 1. Device structures of the magneto-plasmonic metasurface.** (a) Device schematics and operation mechanism. An out-of-plane magnetic field (PMOKE configuration) modulates the far field extrinsic chirality under oblique incidence conditions. (b) Cross sectional image of the device characterized by TEM showing the structure and thicknesses of each layer. (c) Surface morphology of the device measured by SEM showing hexagonal periodic nanohole structures in the Au thin film fabricated by PS sphere self-assembly.

**Device structures of the magneto-plasmonic metasurface.** Figure 1a shows the device structure and operation mechanism. From the top surface to the bottom. The device is composed of a thin Au membrane fabricated on top of a Ce:YIG/YIG/SiO$_2$/TiN multilayer, deposited on top of a SiO$_2$ substrate The operation mechanism can be understood as follows. Due to a centrosymmetric hexagonal structure of the nanoholes, the device shows extrinsic chirality when circular polarized light is obliquely incident on the metasurface.[12] This optical chirality is originated from an asymmetric electromagnetic field distribution in the nanohole structure when plasmonic resonant modes are excited.[12] On the other hand, and due to the MO activity of the Ce:YIG layer, a magnetic field applied along the surface normal induces a difference in the optical response of the system for left and right circularly polarized light, allowing magnetic field modulation of the CD signal. Moreover, due to the metal (Au)-dielectric-metal (TiN) (MIM) cavity structure of the multilayer film stack, this magnetic modulation is enhanced by the strong localization of the electromagnetic field in the Ce:YIG layer at the cavity mode. Note that the YIG layer acting as a seed layer for Ce:YIG crystallization also shows magneto-optical effect. But the amplitude is about one order of magnitude lower than Ce:YIG,[29] which induces minor contribution to the CD modulation. For device fabrication, the Ce:YIG and YIG thin films are deposited by pulsed laser deposition (PLD), while TiN, SiO$_2$ and Au thin films are deposited by sputtering (see methods). The perforated Au thin film is fabricated by PS sphere self-assembly, thermal evaporation and a lift-off process (see Supporting Information Figure S1). The crystal structures are characterized by X-ray diffraction (XRD), confirming the formation of garnet phases in YIG and Ce:YIG[29] (see Supporting Information Figure S2a). The thickness of each thin film layer is TiN (129 nm)/SiO$_2$ (5 nm)/YIG (49 nm)/Ce:YIG (47 nm)/Au (51 nm), as characterized by cross-sectional transmission electron microscopy (TEM) shown in Figure 1b. Energy dispersive spectroscopy (EDS) characterization in TEM shows sharp interfaces and little inter-diffusion between layers during the fabrication process (see Supporting Information Figure S2c). Figure 1c shows the surface morphology of the perforated Au thin film measured by scanning electron microscopy (SEM). A uniform hexagonal periodic structure is observed. The period and radius of the nanoholes are 540 nm and 185 nm, respectively, which are also confirmed by atomic force microscopy (AFM) images in Supporting Information Figure S2b.

**Extrinsic chiroptical properties of the metasurface.** We firstly measure the CD spectrum of the devices under oblique incidence and zero applied magnetic fields using the Mueller matrix method on a spectroscopic ellipsometer, as shown in Figure 2a. Here, The CD is calculated by:[30]

$$CD = \Delta A = A_{RCP} - A_{LCP} = \log_{10}(\frac{1-M_{14}}{1+M_{14}}) \qquad (1)$$

where $M_{14}$ is the matrix element of the normalized Mueller matrix,[30] $A_{RCP}$ and $A_{LCP}$ are the absorbance of the right and left circular polarized light respectively. $1-M_{14}$ and $1+M_{14}$ represent the reflection for left and right circularly polarized light, respectively.[30] Because our structure show no transmittance at the tested wavelength range due to the thick TiN layer, the absorbance is calculated by $A = -\log_{10}(R)$, where $R$ is the reflectance of the device. Here, we use the equation 33(log(1-$M_{14}$)-log(1+$M_{14}$)) to convert the CD unit to degrees.[19] For normal incidence, the CD spectra is measured by the free-space CD characterization set-up. With increasing the incident angle from 45° to 60°, the CD peak is redshifted to longer wavelengths. A maximum CD of 1.5° is observed at around 960 nm wavelength with 60° incident angle. The redshift is due to coupling of the cavity mode and the surface plasmon mode of Au/CeYIG (1,0) at high incident angles, where (1,0) stands for the diffraction order of the hexagonal nanohole grating structure, as shown in Supporting Information Figure S3a. For the surface plasmon mode (SPP) of a hexagonal periodic structure, the wavevector of the surface plasmon mode can be expressed as:[31]

$$\vec{k}_0 \sqrt{\frac{\varepsilon_m \varepsilon_d}{\varepsilon_m + \varepsilon_d}} = \vec{k}_0 \sin\theta + i\frac{4\pi}{\sqrt{3}P}\vec{b}_1 + j\frac{4\pi}{\sqrt{3}P}\vec{b}_2 \qquad (2)$$

where $\vec{k}_0$ is free space wave-vector, $P$ is the grating period, $i$ and $j$ are the diffraction orders of the grating, $\theta$ is the incident angle, $\vec{b}_1$ and $\vec{b}_2$ are unit vector of reciprocal hexagonal lattice, $\varepsilon_m$ and $\varepsilon_d$ are the dielectric constants of the metal and dielectric layer respectively. The CD value is proportion to the incident angles, which is due to the large electric field asymmetry at high incident angles.[14,32] The calculated CD spectra are shown in Figure 2b, which is consistent with experimental results. The spectrum is redshifted with increasing the incident angle, agreeing with experimental results. The experimental CD spectrum shows wider peak widths compared to the simulation results, possibly due to an underestimation of the optical loss in polycrystalline Au and Ce:YIG thin films. As stated before, the large extrinsic optical chirality is due to an asymmetric distribution of the electromagnetic field under oblique incidence. To show this, we simulated the near-field distribution of the normalized electric field for 0° and 60° incident angles at 950 nm wavelength, as shown in Figure 2c-2f. For perpendicular incidence, the electric field distribution of RCP and LCP incident light are almost identical, leading to zero CD value. However, for 60° incidence, the normalized electric fields of RCP and LCP incident light are different, leading to large CD signals in the far-field. In order to confirm the correctness of the Mueller matrix method, we also measure the CD spectrum by a home-built free-space CD characterization set-up (see Supporting Information Figure S4). In this case, the CD measured by free space optics is defined as:

$$CD = A_{RCP} - A_{LCP} \qquad (3)$$

where $A_{RCP}$ and $A_{LCP}$ are the absorbance of RCP and LCP incident light, respectively. The results for both experiments agree with each other. The experimental absorption spectra of right (RCP) and left (LCP) circular polarized light with incident angles changing from 45° to 60° are also shown in Supporting Information Figure S6a, indicating a cavity mode excitation at around 950 nm wavelength.

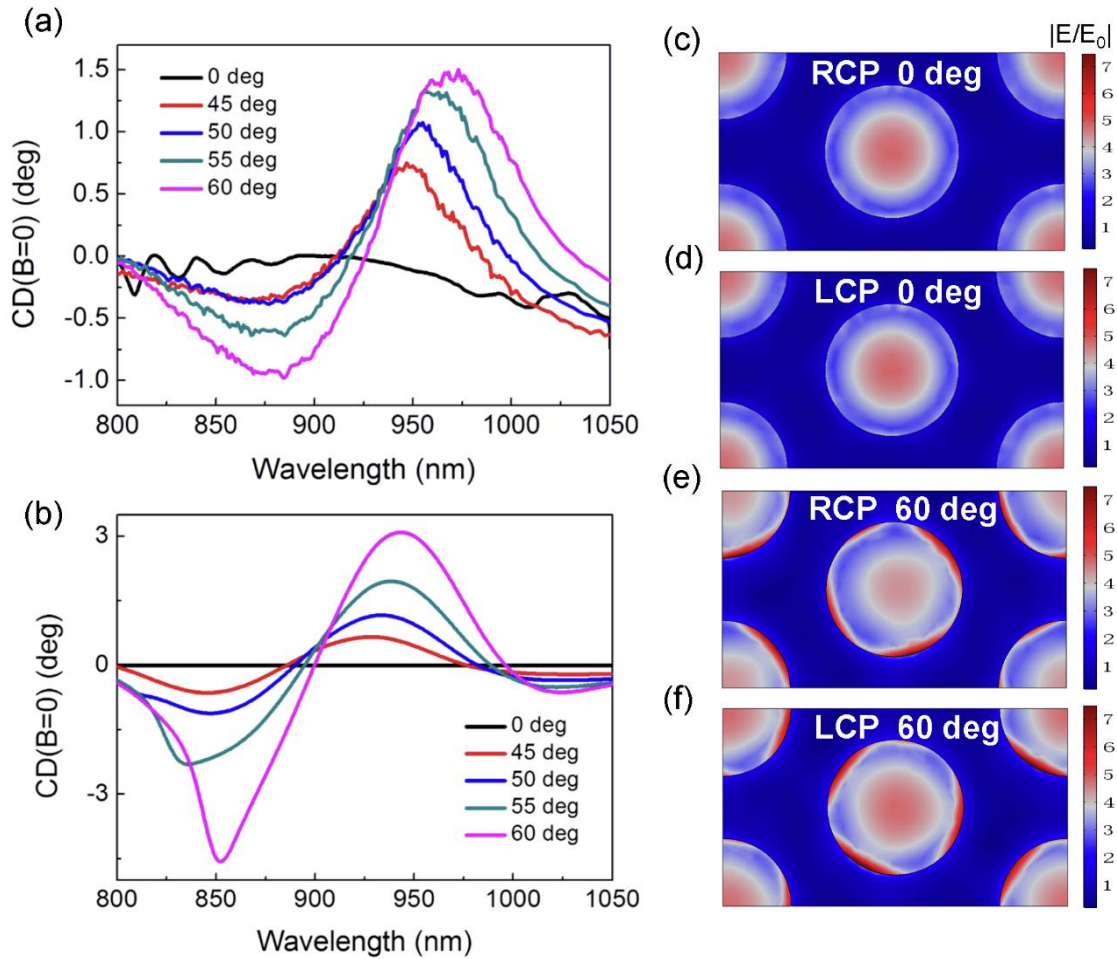

**Figure 2. Experimental and simulated CD spectrum at zero applied magnetic field.** (a) Experimental CD spectra of the metasurface device with different incident angles. (b) Simulated CD spectra of the metasurface device with different incident angles. (c-f) Electric field distribution (normalized to free space field intensity of the incident light) at 950 nm wavelength for RCP and LCP light at 0° and 60° incidence angles. The field is plotted at the Au and Ce:YIG thin film interface.

**Magnetic field tunable far field optical chirality.** Next, we characterized the effect of applying a magnetic field to switch the extrinsic chirality. Upon the application of a magnetic field, the CD signal can be expressed by[33,34]

$$CD(B) = CD(B=0) + MCD \qquad (4)$$

where *CD(B=0)* is the extrinsic chirality of the nanohole structure under zero applied magnetic field, *MCD* is the magnetic circular dichroism originated from Ce:YIG. Note the index change of Ce:YIG under applied magnetic fields also lead to modulation of the near field electromagnetic field distribution,[26] which modulates the extrinsic chirality and is included in the MCD term. Thanks to the field localization in Ce:YIG, the far field CD is strongly influenced by the magneto-optical effect of Ce:YIG, therefore leading to a significant modulation of the CD signal by applied magnetic fields. Figure 3a shows the magnetic field modulation of the CD measured by the Mueller matrix method with different out-of-plane applied magnetic fields for the 45° incidence condition. To apply magnetic fields, we placed a small permanent magnet (NdFeB) behind the sample with the surface magnetic field up to 3.1 kOe as measured by a gauss meter. By flipping the polarity of the permanent magnet, negative magnetic fields can also be applied. By switching the magnetic field polarity, the CD signal at 950 nm wavelength is switched from -0.65° to +1.9°. To demonstrate continuously tunable CD signals by applied magnetic fields, we measured the CD spectra by applying different strengths of magnetic fields, as shown in Figure 3b. This is achieved by placing the sample at different distances from the permanent magnet. The CD signal gradually changes from negative to positive with changing the magnetic field from -3.1 kOe to +3.1 kOe, indicating a continuously tunable CD spectrum. In Figure 3c, we plot CD versus the applied magnetic field at 950 nm wavelength. In this field range, the CD signal scales almost linearly with the applied magnetic field both for the positive and negative side, which is consistent with the magneto-optical hysteresis loops shown in Supporting Information Figure S8. The error bar indicates the standard deviation of five measurements. Figure 3d shows the simulated CD spectra at 950 nm wavelength. The simulation results are consistent with the experimental results, with a CD tuning range from -1° to +1.9° at 950 nm wavelength. The difference between experiment and simulation is possibly due to a lower applied magnetic field compared to the saturation magnetic field required for Ce:YIG (the out plane saturation magnetic field is about 4 kOe, as shown in Supporting Information Figure S8). We also simulated and measured the CD spectra at different incident angles and under positive and negative applied magnetic fields, as shown in Supporting Information Figure S9 and Figure S10. As we increase the incident angle the MCD decreases. On the other hand, the *CD(B=0)* signal increases with increasing the incident angles due to a stronger asymmetry of the electromagnetic field distribution, as shown in Figure 2a. Hence, for small incident angles, a lower *CD(B=0)* signal and larger magnetic field modulation amplitude is observed. Whereas for large incident angles, a higher *CD(B=0)* is observed, and a lower modulation amplitude by the magnetic fields is observed.

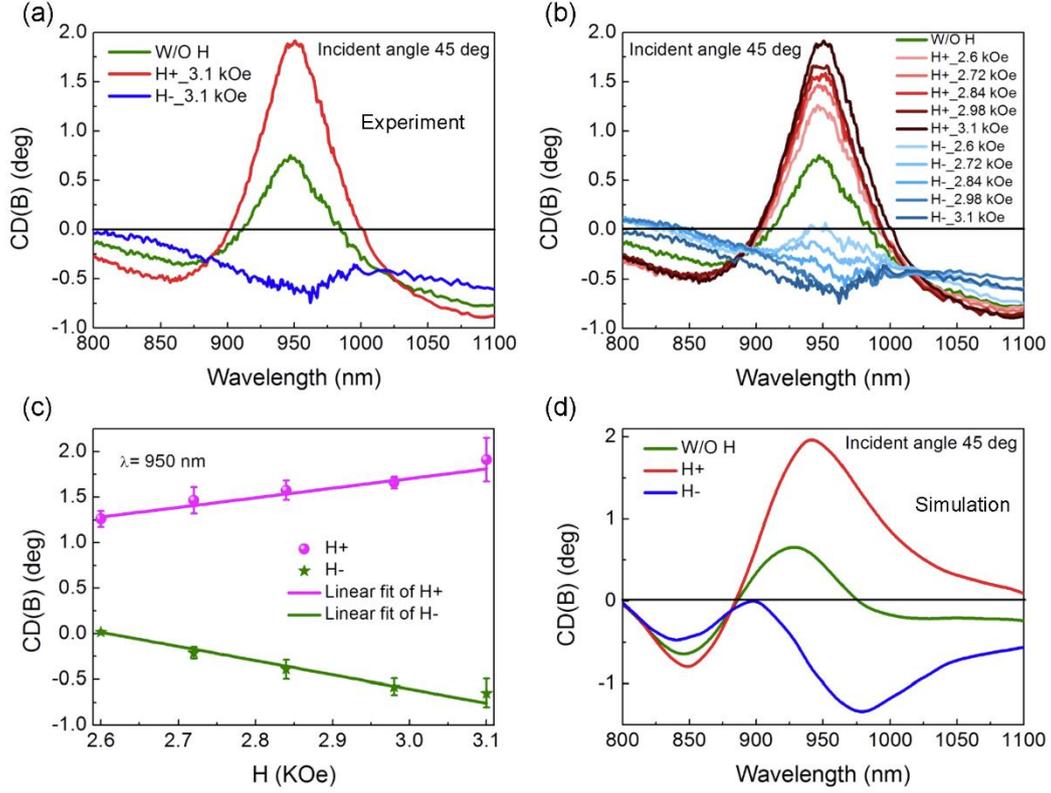

**Figure 3. Switching the extrinsic chirality by applied magnetic fields.** (a) The experimental CD spectra under 0 and 3.1 kOe applied magnetic fields at 45° incidence. (b) The experimental CD spectra for the 45° incidence condition and under applied magnetic fields changing from -3.1 kOe to +3.1 kOe. (c) CD as a function of applied magnetic fields at 950 nm wavelength. The error bar represents the standard deviation of five measurements. (d) The simulated CD spectra with an incident angle of 45° under zero, positive, and negative saturation magnetic fields.

**Magnetic field tunable near field the optical chirality.** To investigate the modulation of the near field chiroptical response by magnetic fields, we quantify this phenomenon by investigating the optical chirality (OC) parameter modulated by magnetic fields, which is defined as:[10]

$$C = -\frac{\varepsilon_0 \omega}{2C_0} \text{Im}(E^* \cdot B) \qquad (5)$$

where $\varepsilon_0$ is the permittivity of vacuum, $\omega$ is the angular frequency, $E^*$ and $B$ are the complex conjugate of the electric field and the magnetic field, $C_0$ is the modulus of the optical chirality of circularly polarized wave (left or right) in the vacuum ($C_0=\varepsilon_0\omega/(2c)*E_0^2$). In Figure 4a, we simulated the OC distribution of RCP light at 950 nm wavelength where the cavity mode was excited and the largest CD is observed. To be consistent with far field CD characterization experiments, the incident angle was set as 45°. The OC is investigated at the Au/Ce:YIG interface in our simulations, where the cavity mode is the strongest in the Ce:YIG layer. An OC up to 10 is observed with the largest amplitude concentrated in the middle and edge of the holes, consistent with the electric field distribution of a cavity mode. Compared to the RCP light, the optical chirality is reversed in sign and comparable in amplitude for the LCP incident light, as shown in Figure 4b. The difference of the OC between the RCP and LCP incident light contribute to the high extrinsic chirality of the achiral structure. As shown in Figure 4c, the difference of the OC is in the range of -20~20, which is comparable to intrinsic chiroptical structures such as gammadions.[1] Magnetic field modulation of the superchiral field can be simulated by considering the permittivity tensor of Ce:YIG as a function of the material magnetization along the out-of-plane (z) direction:

$$\tilde{\varepsilon}=\begin{pmatrix} n_0^2 & -i\gamma\frac{M}{M_s} & 0 \\ i\gamma\frac{M}{M_s} & n_0^2 & 0 \\ 0 & 0 & n_0^2 \end{pmatrix} \qquad (6)$$

where $n_0$ is the complex refractive index of Ce:YIG without applying magnetic fields, $\gamma$ is the complex off-diagonal permittivity tensor element when Ce:YIG is magnetized to saturation. $M_s$ is the saturation magnetization of Ce:YIG, and $M$ ($|M| \leq M_s$) is the magnetic moment of Ce:YIG, which is a function of applied magnetic field $H$ determined by the magnetization hysteresis. Figure 4d and 4e shows the simulated OC difference between $+M_s$ and $-M_s$ magnetization condition of Ce:YIG for the RCP and LCP light respectively. The modulation amplitude reaches ±0.6, which is about an order of magnitude higher compared to Au/Co gammadions nanostructures[10] ($C_{R/LCP\uparrow/\downarrow}$ represent the OC of right/left circular light with positive/negative magnetic fields. $\Delta C_{\uparrow\downarrow}$ is equal to ($C_{RCP\uparrow}$-$C_{LCP\uparrow}$)-($C_{RCP\downarrow}$-$C_{LCP\downarrow}$)). Therefore, the magnetic field modulation amplitude of the local OC ($2 \times (C_\uparrow-C_\downarrow)/(C_\uparrow+C_\downarrow)$) is around 6%. Unlike the non-magnetic component (Figure 4a, b, c) where a large OC is observed at the hole center, the highest amplitude of magnetic field modulation of the OC is observed at the edge of the hole, due to a larger field intensity in Ce:YIG at the edge. Finally, in Figure 4f, we also simulated the magnetic component of the extrinsic chirality, which was also about an order of magnitude higher compared to previous reports.[10,26] Therefore, the large tuning range of the OC by magnetic fields makes our device potentially useful for active chiral photonic device applications.

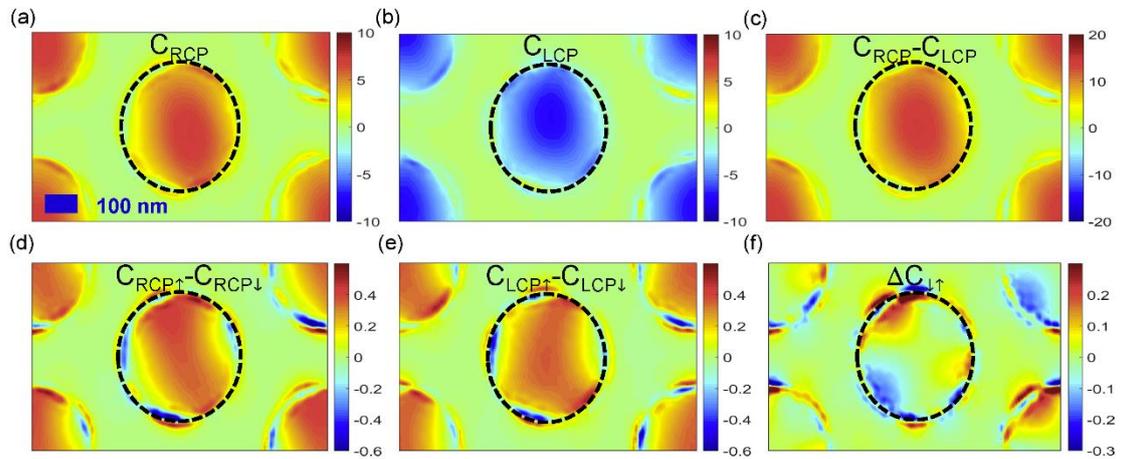

**Figure 4. Optical chirality of the metasurfaces.** (a) OC distribution of metasurface with RCP incident light at 45° incident angle. (b) OC distribution of metasurface with LCP incident light at 45° incident angle. (c) The difference of the OC between RCP and LCP incident light. (d) The difference of the OC of RCP incident light under applied saturation magnetization field along opposite directions. (e) The difference of the OC of the LCP incident light under applied saturation

magnetization field along opposite directions. (f) Magnetic field modulation of the OC difference. The dashed circles outline the boarder of one perforated nanohole. All simulations are performed at the interface of Au nanohole and Ce: YIG layer.

**Magnetic field tunable chiral images.** To visualize magnetic field switching of the optical chirality of the metasurface, we fabricate a school badge pattern of "UESTC" on the Ce:YIG/YIG/SiO$_2$/TiN multilayer films in a 2 mm by 2 mm area using photolithography and self-assembly of PS spheres. Figure 5a shows the device image taken by lenses and an infrared CCD for the RCP incident light with the wavelength of 950 nm and incident angle of 45°. The reflectivity was recorded by an image capture card as greyscale values and replotted in MATLAB. The blue regions with lower reflectivity are patterned with Au nanohole structures, *i.e.* the same structure as Figure 1a, as shown in the enlarged view. Whereas the golden regions with higher reflectivity are the bare Ce:YIG/YIG/SiO$_2$/TiN multilayer thin films. Figure 5b shows the reflectivity spectra of the hole area and bare multilayer thin films. In Figure 5c, we extract the CD image from the RCP and LCP light images, which is plotted by calculating reflectance circular dichroism (RCD=($R_{RCP}$-$R_{LCP}$)/($R_{RCP}$+$R_{LCP}$)) at each pixel point.[35,36] Where the $R_{RCP}$ and $R_{LCP}$ represent the reflectivity of right and left circular polarization incident light, respectively. Upon applying upward or downward magnetic fields of 3.1 kOe, we observe obvious RCD sign reversal in the whole image at the metasurface regions indicated by color changing from yellow (positive RCD) to blue (negative RCD). The CD modulation amplitude reaches 6° in most metasurface regions, which is comparable to the single point measurement results. This result demonstrates the possibility to fabricate large scale active chiroptical metasurfaces using magneto-optical materials, which is promising for imaging and sensing applications.

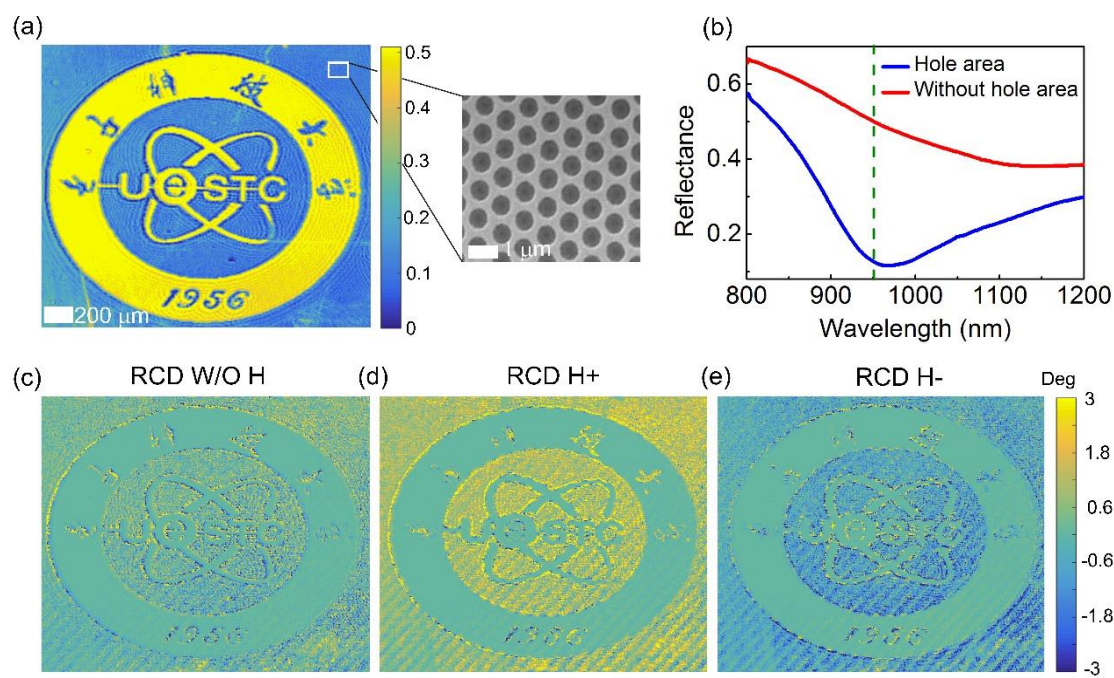

**Figure 5. Magnetic field modulation of the chiral image of the "UESTC" school badge pattern.** (a) The device image under RCP incident light measured by a CCD and replotted in greyscale reflectivity values. The blue regions are metasurface nanostructures as shown in the zoom-in image. (b) Reflectivity spectrum on and off the metasurface. The green dash line indicates the wavelength used for imaging. (c) RCD image without applied magnetic fields. (d) RCD image with out-of-plane applied magnetic field of +3.1 kOe. (e) RCD image with out-of-plane applied magnetic field of -3.1 kOe.

## CONCLUSIONS

In summary, we report switching of the extrinsic optical chirality in magneto-optical metasurfaces using magnetic fields. Thanks to the strong cavity mode excited in low loss magneto-optical oxide thin films, we demonstrate a sign reversal and large tuning range of the far-field CD spectrum in magnetoplasmonic metasurfaces by applied magnetic fields from -0.65° to +1.9°, accompanied with a large modulation amplitude of the near-field optical chirality up to 0.6, which is one order higher compared to previous reports. We also demonstrate large scale fabrication capability of such tunable metasurfaces by presenting a millimeter scale metasurface device and magnetic field tunable chiral images. Our results demonstrate a promising potential of magnetic field tunable chiral metasurfaces based on low loss magnetic oxides for integrated polarization control, sensing and display applications.

## METHODS

**Sample Fabrication.** The multilayer thin film stack of 129 nm TiN/5 nm $SiO_2$/49 nm YIG/47 nm CeYIG/51 nm Au was deposited on silica substrates using pulsed laser deposition (PLD) and magnetron sputtering. The bottom TiN film was first deposited by PLD (TSST) equipped with a 248 nm KrF excimer laser in 0.5 Pa nitrogen ambient at 800 °C. The base pressure before deposition was $1\times10^{-6}$ Pa. The laser fluence was 3 $J/cm^2$. After deposition, the sample was cooled down in flowing nitrogen at a rate of 5 °C/min to room temperature. Then the sample was transferred to a sputtering chamber with the base pressure of $5\times10^{-4}$ Pa. A thin layer of $SiO_2$ was sputtered onto the TiN film to prevent TiN oxidation and promote YIG and Ce:YIG crystallization. A 49 nm thick YIG thin film was deposited on the $SiO_2$ layer at room temperature by PLD and rapid thermal annealed in nitrogen ambient at 860 °C for 5 min to crystallize. The YIG acted as a seed layer for Ce:YIG thin film to crystallize which showed a higher magneto-optical effect. After YIG thin film crystallization, the Ce:YIG thin film was deposited at room temperature on YIG, and rapid thermal annealed in nitrogen at 800 °C for 3 min.

The perforated Au thin film was fabricated by polystyrene (PS) sphere self-assembly and lift-off technology. Firstly, PS sphere powders were dispersed in a mixture solution of water and ethanol (volume ratio 1:1) to prepare the PS sphere dispersed solution with a concentration of 15 wt %. Then, the PS sphere dispersed solution was uniformly dispersed under ultrasonic cleaning for 1 hour. To assemble the PS spheres, we slowly drop the PS sphere dispersed solution onto a water surface using a pipette. After that, a few drops of 0.1 wt % sodium dodecyl sulfate (SDS) were added to promote a dense self-assembly of PS spheres. After the PS sphere was self-assembled into a dense packed hexagonal lattice on the still water surface, we transferred the self-assembled PS spheres from water to sample surface. The multilayer film sample was placed into the water, underneath the PS sphere layer. Afterwards, the water was slowly pumped out using a peristaltic pump. After pumping the water, the PS sphere was transferred to the surface of the films. Then we dried the sample in ambient conditions for 2 hours followed by baking the sample for 1 min on a hot plate at 100 ºC. Then, the PS sphere was oxidized to smaller diameters from 547 nm to 360 nm in oxygen plasma (80 W, 250 s). Au thin film was then deposited on the sample surface by thermal evaporation (Leybold UNIVEX250) at room temperature with a base pressure of $1\times10^{-4}$ Pa. The perforated Au thin film was then fabricated by rinsing off the PS sphere, by soaking into a methylbenzene solution with ultrasonic cleaning for 3 min, and then rinsed in deionized water for 1 min.

**Numerical Simulations.** Commercial numerical software (COMSOL MULTIPHYSICS®) based on finite element method was used to simulate the far and near field response of the device. In our model, we used a hexagonal nanohole structure unit with periodic boundary conditions. The circular polarization light was defined in periodic port. A perfectly matched layer (PML) was set below the TiN layer along the propagation direction to absorb the scattered light. The reflectance spectra were calculated by the S-parameters in COMSOL. The OC was simulated at the interface between Au and Ce:YIG layer. For permittivity of Au and TiN, we use the Drude model[37] and Drude-Lorentz dispersion model[38] to fit the permittivity. The refractive index of YIG is obtained from our experimental data measured by the spectroscopic ellipsometer. The permittivity tensor of Ce:YIG is obtained from Ref 39,40.

**Measurement Procedure.** The reflectance and CD spectra were measured by a home-built free space CD characterization set-up as shown in Supporting Information Figure S4. The characterization set-up includes a supercontinuum laser (NKT Photonics), a 1/4 wave plate, a compensated full-wave liquid crystal retarder (Thorlabs, LCC1413-B) and a rotation table. The right and left circular polarization were controlled by the LCC1413-B by applying different voltages to cause birefringence of the liquid crystal. The reflection spectrum as a function of wavelength was collected at 5° angle intervals. The reflectance as a function of the wavelength and incident angles was plotted, indicating the dispersion relation of the sample. The CD spectra were also characterized by measuring the Mueller matrices using a spectroscopic ellipsometer (J. Woollam RC2) in a wavelength range from 230 to 1690 nm. Kerr spectra and Kerr Loops were obtained at nearly normal incidence with a magnetic field, generated by an electromagnet, applied perpendicular to the sample surface (see supplementary information). To measure the magnetic field modulation of CD, we used a NdFeB permanent magnet with the surface magnetic field up to 3.1 kOe confirmed by a Gauss meter. For chiral imaging, we replaced the photodetector with a CCD. A black silicon CCD with response wavelength up to 1310 nm and 400,000 pixels was used to detect the chiral images.

**Author Contributions**

J. Q. conceived the idea. J. Q., X. L., and T. T. performed the theoretical calculation and numerical simulations. J. Q., T. K., L. N., R. Y., H. W., and G. A. fabricated the samples and performed experiments. H. F., C. L., H. W, and Y. L. helped with the theoretical interpretation. L. D., G. A., and L. B. supervised the project. All authors discussed the results and prepared the paper.

**Notes**

The authors declare no conflict of interest.

**ACKNOWLEDGEMENTS**

Ministry of Science and Technology of China MOST (2016YFA0300802), International collaboration platform supported by Ministry of Education ("111 Project" of China) (B18011) are acknowledged.